# Typhoon Tracks Regulated by Feedbacks of Fine-Scale Clouds to Environment

Haoran Zhao[1], Shaoqing Zhang[1,*], Yang Gao[2,*], Lixin Wu[1,3,*], Yihan Cao[1], Wenju Cai[1,3], Bin Wang4, L. Ruby Leung5, Zebin Lu[1], Zhong Zhong6, Xiaolin Yu[1], Mingkui Li[1], & Chenyu Zhu[1]

**Affiliations:**

[1]Frontiers Science Center for Deep Ocean Multispheres and Earth System, and Key Laboratory of Physical Oceanography, Ministry of Education, the College of Oceanic and Atmospheric Sciences, Ocean University of China, Qingdao, China

[2]Frontiers Science Center for Deep Ocean Multispheres and Earth System, and Key Laboratory of Marine Environmental Science and Ecology, Ministry of Education, Ocean University of China, Qingdao, 266100, China

[3]Laoshan Laboratory, Qingdao, China

[4]Department of Atmospheric Sciences and International Pacific Research Center, University of Hawaii at Manoa, Honolulu, HI, USA

[5]Atmospheric Sciences and Global Change Division, Pacific Northwest National Laboratory, Richland, WA, 99354, USA

[6]College of Meteorology and Oceanography, National University of Defense Technology, Changsha, China

(To be submitted to *Nature*)

*Corresponding author. Email: szhang@ouc.edu.cn; yanggao@ouc.edu.cn; lxwu@ouc.edu.cn

**Accurate tropical cyclone (TC) track prediction is crucial for mitigating TC's catastrophic impacts on human life and the environment. Despite decades of research on tropical cyclone (TC) track prediction, large errors known as track forecast busts (TFBs) occur frequently, and their causes remain poorly understood. Here, we examine a few dozens of TCs using a unique TC downscaling strategy that can quantitatively assess the sensitivity of TC track on the strength of feedbacks of fine-scale clouds to environment. We show that as TFBs have a weaker environmental steering that favors scattering cumulonimbus clouds, capturing asymmetric distribution of planetary vorticity advection induced by such fine-scale clouds corrects TFBs by 60%. Our clear identification of such important TC track predictability source promises continuous improvement of TC track prediction as finer-scale TC clouds and their interactions with environment are better resolved as model larger-scale behaviors have improved.**

Tropical cyclones (TCs), strong ones in the western North Pacific called typhoons, are intense circular storms originating in the tropical ocean, frequently causing severe disasters with tremendous life and property losses[1-3] and exerting substantial impacts on climate[4]. Accurate TC track forecast is crucial for mitigating TC's catastrophes by enabling precisely-positioned early warning and reducing uncertainties in other elements (e.g. intensity) associated with local conditions[5]. During the past few decades, prediction of TC tracks has progressed with advancement of numerical weather prediction (NWP) and improvement of observational network[6-8]. However, TC track forecast busts (TFBs) with large track prediction errors occur frequently in the current NWP operational models[9-12]. At present, TFBs are a major obstacle to further advancing TC simulation and prediction for weather and climate studies[3]. Recently, powerful artificial intelligence (AI) forecasting methods have greatly improved general weather prediction[13-15], including TC track forecasts[14], but without clear understanding of TFB predictability, such data-driven approach has substantial limitation to further enhance the prediction skill of TFBs as extreme events[14,16]. There is an urgent need to clearly identify the predictability sources of TFBs so as to further guide the development of global NWP models and help AI advances.

TC motion is primarily steered by the environmental flows, together with the *β*-effect caused by the latitudinal variation of the Coriolis parameter and diabatic heating associated with TC clouds[17-19]. While TC tracks in most of cases are governed by environmental flows, TFBs case studies suggest that large track errors are associated with the structural bias of TC clouds, partly due to the systematic errors caused by deficiencies in physics of clouds[20-22] and other model imperfections[3,23,24]. Specifically, uncertainties in TC clouds can translate into the environment and differ it through the *β*-drift produced by the advection of planetary vorticity by the storm-scale cyclonic circulation[19], leading to biases in TC tracks. Case studies also indicate that the environment and *β*-drift can exert a significant impact on TC clouds, leading to the occurrence of concentric eyewalls, eyewall replacement, and intensity changes[25-27]. However, the existence of uncertainties in the strength and phase of convective asymmetries in numerically-predicted TC evolution, how the interior structure of TC clouds, *β*-drift and environment work together to control the track of TFBs is still unclear and remains an open issue.

One method commonly used to reduce uncertainties of TC clouds in numerical simulations is through convection-permitting high-resolution modeling, to better simulate the fine-scale structure of TC clouds in the eyewall and rainbands[28-30]. However, uncertainty reduction in TC clouds does not necessarily lead to systematical improvement in simulating the movement of TFBs[23,29,31]. On the one hand, assessments of multiple TCs including small track errors (STEs) and TFBs are unable to sort out the predictability sources of TFBs through simulations at various resolutions[31,32], without deepening detection of the feature of TFBs. On the other hand, without an effective experimenting approach that can quantitatively assess the linkage among TC clouds, *β*-drift and large-scale environment as well as their roles in controlling TC track, it is very difficult to establish a clear picture of track predictability of TFBs thus significantly reducing their occurrences.

Here, we first design a unique TC downscaling strategy that can test the change of TC track as the influencing strength of TC clouds as fine as resolved by 1 km grid-spacing on $\beta$-drift and environment changes. Then, we examine a few dozens of typhoons from 2018 to 2021 and detect the distinctive features of TFBs versus STEs. We discover that the feedback of fine-scale clouds to the environmental flow is the key source of track predictability of TFBs. The diabatic heating/cooling of asymmetric distribution fine-scale TC clouds first alters the local $\beta$-drift through linear beta term and nonlinear advection of the asymmetric momentum anomalies by the symmetric vortex flow, which feeds to the environmental flow. The environmental flow further reconfigures the TC clouds which in turn modulate the environmental flow. Such iterative interactions eventually establish a new TC steering that controls the correct track for TFBs.

**Model evidence of feedbacks of fine-scale TC clouds to environment reducing TFBs errors**

As destructive TCs have become more frequent in the last two decades (**Supplementary Fig. 1**), we focus on typhoons (maximum sustained wind, MSW ≥ 32.7 m/s)[33,34] in the West Pacific during 2018-2021 (**Supplementary Tab. 1**). We first assess the prediction statistics of the 27v9 (27 km resolution atmosphere coupled with 9 km resolution ocean) Asia Pacific Regional Coupled Prediction (AP-RCP) model[28] (see **Methods**). Out of 93 TCs observed in the Joint Typhoon Warning Center (JTWC), 46 are selected for this study as typhoon or higher categories with lifetime ≥ 3 days (in which 43 typhoons have lifetime ≥ 5 days). The track prediction error of AP-RCP is representative of typical numerical weather prediction although the 5$^{th}$-day mean track error (378 km) of the 43 typhoons is larger than the mean error (~300 km) of TCs over all categories[8]. More specifically, 8 TFBs, Shanshan, Soulik, Yutu and Man-Yi (2018), Krosa and Kammuri (2019), and In-Fa and Rai (2021) have much larger track errors (> 500 km) than the mean error of the 5$^{th}$-day prediction, being selected as the target of this study. These 8 TFBs include 3 with sudden moving-direction change (Kammuri, Soulik and In-Fa), 2 with slow moving-direction change (Man-Yi and Rai), 2 without significant moving-direction change (Shanshan and Yutu), and 1 with a complex zigzag moving-path (Krosa) (**Fig. 1A**). Focusing on these 8 TFBs with multiple path features, we use multi-nested downscaling simulations to examine the mechanism of predictability of TFBs.

We take the AP-RCP atmospheric component with moving-TC nests to perform downscaling simulations, which include the base simulation at 27 km resolution and a set of multi-layer, moving nest simulation down to 9, 3 and 1 km resolutions (see Methods). With the same initial and boundary information as used for the 27v9 AP-RCP predictions, we first re-run the 8 TFBs with the multi-layer downscaling using one-way and two-way nesting, in which the two-way allows fine-scale information being sent back to the coarser domain, i.e. with feedbacks, while the one-way does not, i.e. without feedback. As the grid resolution increases, the simulations show general improvement in reproducing the observed TC tracks (**Fig. 1B**) although the improvement at 9 km is subtle for both one-way and two-way nesting. Notably, two-way nesting at 3 km substantially improves the $5^{th}$-day track prediction, with an error reduction that is almost 4 times that at 9 km when both are compared to the base 27 km without nesting, while one-way nesting at 3 km shows more modest skill similar to 9 km. More specifically, compared to the base 27 km resolution predictions, the two-way nested 3 km resolution predictions reduce the $5^{th}$-day ($3^{rd}$-day) track error by 59% (26%). Further increasing resolution from 3 km to 1 km has negligible effect on the track prediction. The large difference between one-way and two-way nestings for each of the 8 TFBs (**Fig. 1C**) suggests that the fine-scale TC-environment interactions resolved at 3 km are critically important for improving track prediction of TFBs, although further increasing model resolution can continuously improve TC intensity (**Extended Data Fig. 1**) as revealed in previous studies[23,29,30].

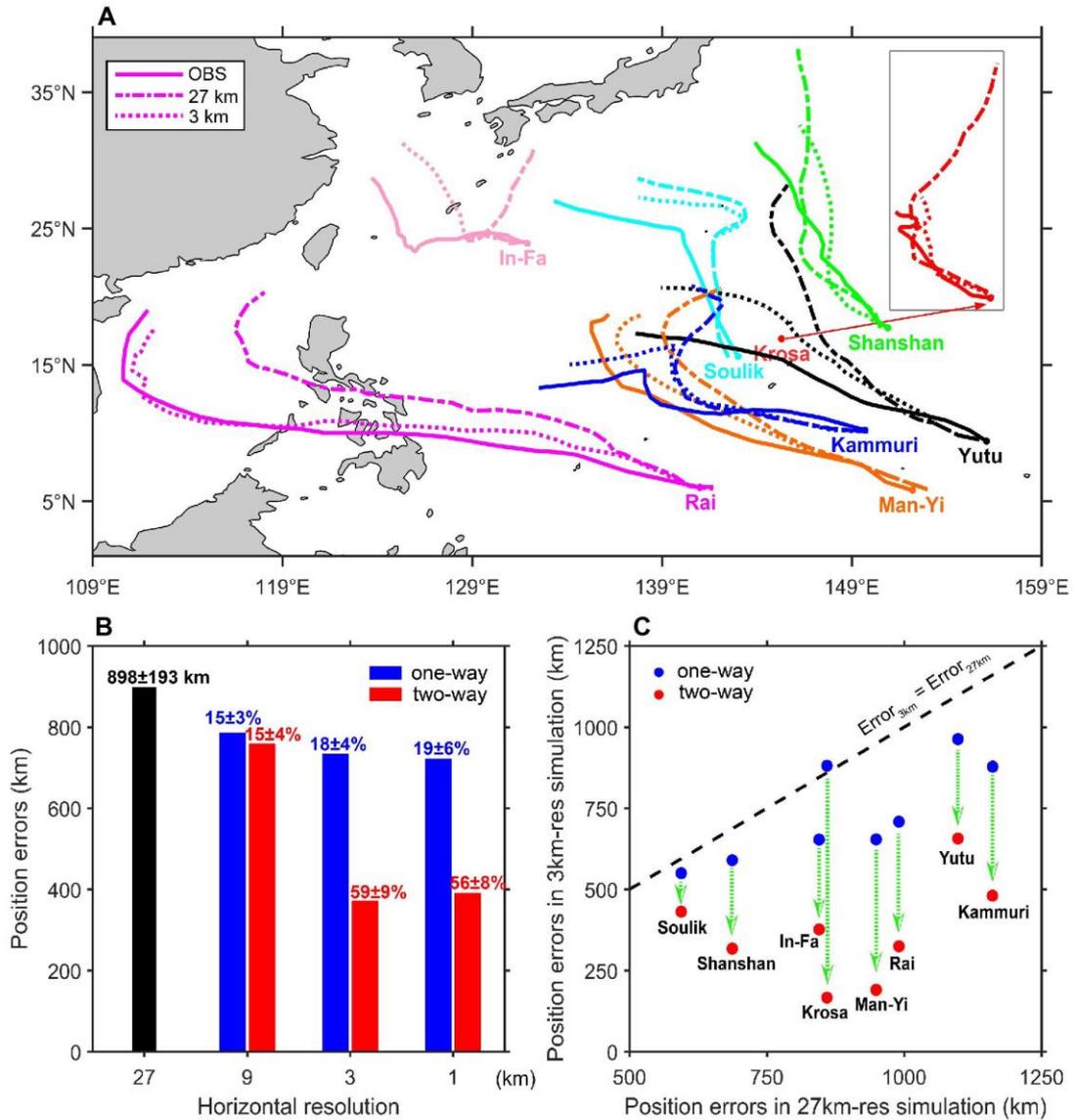

**Fig. 1 | Model evidence of feedbacks of fine-scale TC clouds to environment reducing track errors of TFBs. A,** 27 (dashed-dotted) and 3 (dotted) km resolution Weather Forecast Research model simulated and observed (solid) tracks of 8 track forecast busts (TFBs) Typhoons, Shanshan (1813), Soulik (1819), Yutu (1826), Man-Yi (1828), Krosa (1910), Kammuri (1928), In-Fa (2106) and Rai (2122). **B,** The mean absolute errors of the 5$^{th}$-day positions (km) of 8 TFBs with one-way (without feedback to mother-domain) (blue) and two-way (with feedback to mother-domain) (red) downscalings started from their onsets. The numbers above the colored bars are the 5$^{th}$-day position errors of the base 27km-res prediction (black) or error reduction percentages of finer-res downscalings (color) from the base 27km-res case, and their standard deviation in 8 cases. **C,** Scatterplots of position errors of individual typhoons (dots) for the 5$^{th}$-day predictions in base 27km-res simulation (x-axis) vs. 3km-res (y-axis) one-way (blue) and two-way (red) downscalings, with green arrows showing error reduction. The feedback of fine-scale TC clouds to environment plays a key role in reducing track errors of TFBs.

Considering the critical importance of 3 km resolution for track simulation and much higher computational cost at 1 km resolution (a factor of 8), we extend our modeling for predicting the other 38 typhoons at 3 km resolution using two-way nesting.

Notably 3 km resolution significantly improves the track and intensity predictions of all 46 typhoons (**Extended Data Fig. 2**), with the mean 5$^{th}$-day position error reduced by 267 km compared to 378 km at 27 km, mainly contributed by TFBs.

**Characteristics of environmental steering and *β*-drift of TFBs**

We select 8 small track error cases (STEs) in the 27v9 AP-RCP system with the smallest 5$^{th}$-day track prediction errors (112 km as the mean) among the 46 typhoons for further analysis (**Supplementary Tab. 1**). Totally different from the 8 TFBs (**Fig. 1A**), the simulated tracks of 8 STEs at 3 km resolution with two-way nesting are very close to those in the base 27 km resolution simulation and observations (**Fig. 2A**). To reveal the distinctive characteristics of environmental flow and *β*-drift in driving TC motion of TFBs, we compare the steering and ventilation flows (VTFs) of TFBs and STEs along with the strength of the western Pacific subtropical high (WPSH) in their base 27 km resolution simulations (**Fig. 2B**) and changes due to feedbacks of 3 km resolution downscalings (**Fig. 2C**). The steering and VTFs are computed as the 850-200 hPa averaged mean wind in the range of 400-700 km from TC center[18] and the 850-200 hPa averaged wavenumber-1[35] (maximum asymmetry) wind in a range of 0-400 km around TC center. The WPSH represents the large-scale environment in the West Pacific area (see **Methods**). The magnitude of steering or VTF change shown in **Fig. 2C** measures changes of the entire vector including its length and direction.

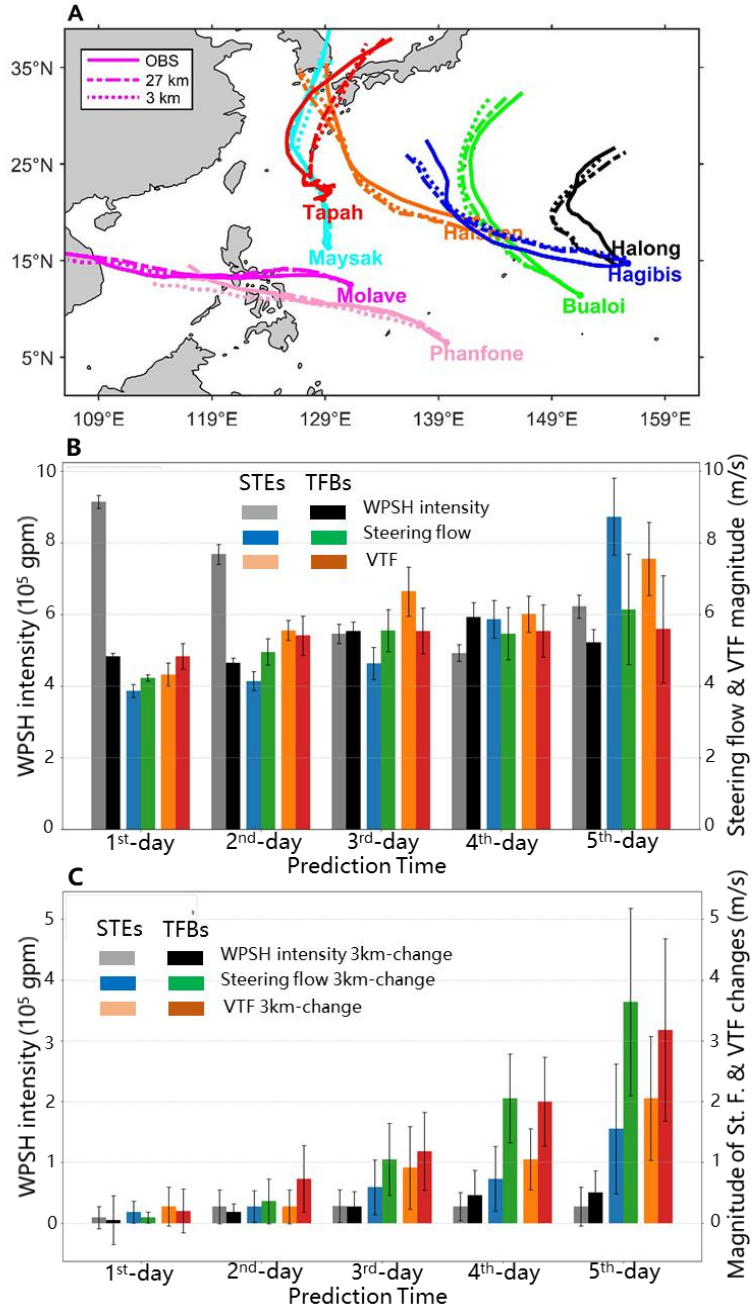

**Fig. 2 | Characteristics of environmental steering and *β*-drift of TFBs. A**, The first 5-day tracks of 8 small track error Typhoons (STEs), Tapah, Hagibis, Bualoi, Halong, Phanfone, Maysak, Haishen and Molave in satellite observations (solid), base 27km-res simulations (dotted-dashed) and 3km-res downscalings (dotted). **B**, Grouped bar charts of first 5-day Western Pacific Subtropical High environment strength (WPSH intensity), TC steering flows and ventilation flows (VTFs) in TFBs vs. STEs as 8-typhoon composite analyses in base 27km-res simulations. **C**, The change of WPSH environment strength and magnitudes of TC steering flow and VTF changes due to feedbacks of 3km-res downscaling from the base 27km-res simulation. The vertical black segments denote the standard deviation in 8 cases. WPSH strength: sum of 500-hPa GHT values $\geq$ 5880 gpm referring to 5870 gpm within the range of 40°×40° centered at each TC[53]; Steering flow: 850-200 hPa averaged mean wind in a range of 400-700 km from TC center[18] ; VTF: 850-200 hPa averaged wavenumber-1[35] (maximum asymmetric) wind in a range of 0-400 km around TC center. The change of TC steering flow or VTF is a vector, the magnitude of which includes the changes of length and direction.

At the early stage, the WPSH of base 27 km simulations in STEs is much stronger than that in TFBs (84% & 50% stronger on the 1$^{st}$-day & 2$^{nd}$-day) but the difference decreases by the prediction time, with almost no difference after the 3$^{rd}$-day. In contrast, as prediction time forwards, the steering and VTFs of STEs in the base 27 km simulations become much stronger than those of TFBs (e.g., 44% stronger for steering and 36% stronger for ventilation on the 5$^{th}$-day). At the same time, the change of steering and VTFs caused by the feedback of 3 km downscaling simulations is much larger in TFBs than STEs, and the difference of base and downscaling simulations increases dramatically with the prediction time. On the 5$^{th}$-day, the changes of steering and VTFs in TFBs are 118% and 60% larger than the changes in STEs. While the environment of TFBs remains weak persistently during the entire 5-day prediction time, the steering of STEs gets strengthened from the strong WPSH large-scale background (see **Extended Data Fig. 3**). While the strengthening of steering flows in the presence of strong large-scale environment for STEs is in stark contrast with the weaker large-scale environment in the base 27 km simulation for TFBs, the change of steering flows brought by 3 km downscaling simulations for TFBs is much larger. These results suggest that for STEs, relative to the strong environmental steering, the interaction between fine-scale clouds and environment is secondary in determining the TC motion. On the contrary, for TFBs, under relatively weak environmental steering, the interaction between fine-scale clouds and environment plays a critical role in determining the TC motion. This is supported by numerical experiments in which artificially strengthening the WPSH for Kammuri reduces the track difference between the base 27 km and 3 km nested simulations, turning Kammuri from a TFB case to an STE case (see **Methods** and **Extended Data Fig. 4**). In the rest of analyses, we will present the physical picture of how the asymmetric planetary vorticity advection induced by fine-scale clouds to fulfill this process.

**Critical importance of resolving cumulonimbus clouds in correcting TFBs' tracks**

To further understand the impact of fine-scale clouds on TFBs' tracks, we first examine the cloud structures of the simulated TCs at different resolutions. We compute the decibel relative to maximum radar reflectivity factor (dBZ) in the eyewall and rainband (the band outside eyewall but clinging it) regions to reflect the characteristics of the TC clouds and analyze the composite of 8 TFBs compared with satellite measurements (**Fig. 3**). When the resolution increases from 27 km to 9, 3 and 1 km, the simulated TC clouds gradually exhibit distinctive characteristics of a spiral cloud system consisting of intensive but asymmetrically-distributive cloud cells as observed by satellite (**Fig. 3A-D**). The simulations at 3 km or higher resolution resolve the fine-scale floccus cloud structures, while the simulated rainband clouds at 9 km or lower resolution cover wide area but the strength at the eyewall is weaker. Wavenumber spectra (see **Methods**) of dBZ simulated at 3 km or higher resolution are close to the spectrum of satellite-measured COT (cloud optical thickness) especially at wavelengths < 10 km, while simulations at 9 km or lower resolution substantially deviate from the observation at wavelengths < 50 km (**Fig. 3E**). In particular, observations show TC eyewall and rainband consisting of intermittent cumulonimbus clouds with characteristic horizontal scales of a few kilometers to 10 km, which provide essential energy for the generation and development of TCs through strong convection[36-38]. Since numerical models cannot permit processes with a length scale less than 2 times of the grid-size[39], the cloud characteristics at 9 km or lower resolution can be attributed to the parameterization of clouds with scales < 20 km. In contrast, the model at 3 km or higher resolution can simulate asymmetric distribution cumulonimbus clouds as shown in the cloud structures of the 8 individual cases (**Supplementary Fig. 3**) much closer to the observed distribution although some discrepancies still exist.

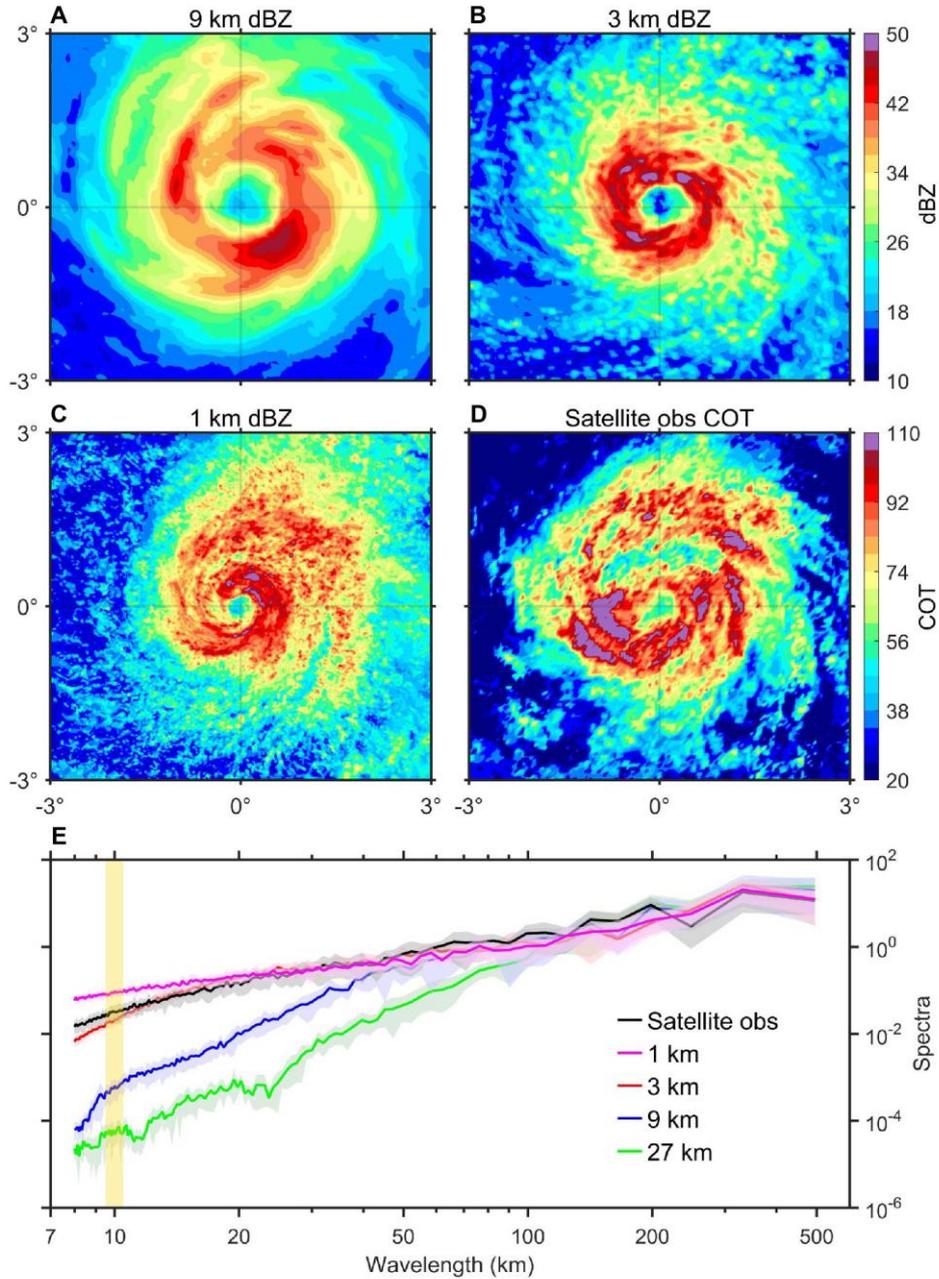

**Fig. 3 | Critical importance of resolving cumulonimbus clouds with ~10 km horizontal scales in correcting TFBs tracks. A-D,** The composite analyses of decibel relative to maximum radar reflectivity factor (dBZ) for 8 TFBs at the 96th-hour predictions with base 27 km resolution, 9, 3 and 1 km resolution two-way downscalings, and corresponding observed cloud optical thickness (COT) from satellite Himawari-8 (in 5 km resolution). **E,** The 8-typhoon composite analysis of dBZ wavenumber spectra in 4 resolution simulations (color) and satellite-observed COT (black). All data are mapped to the 1 km resolution space for spectral analysis. The simulations that resolve cumulonimbus clouds with ~10 km characteristic horizontal scales exhibit a distinctive characteristic of spiral cloud system consisting of intensive small cloud cells as observed by satellite.

To understand the role of cumulonimbus clouds in the development and movement of TFBs, we examine the flows of eddy available potential energy $A_e$ and eddy kinetic energy $K_e$ (see **Supplementary Text 1(a)**) for the 8 TFBs simulated at different resolutions with and without two-way nesting. We analyze the $A_e$ generation [G($A_e$)] and $A_e$ to $K_e$ conversion [C($A_e$, $K_e$)] compared to the distribution of dBZs in the simulations (see **Methods** and **Supplementary Text 1(b)**). Both G($A_e$) and C($A_e$, $K_e$) have consistent behaviors as dBZs converge to the satellite-observed clouds as the TC downscaling resolution increases (**Fig. 3A-D**), displaying the feature of spiral cloud-like that consists of asymmetric distribution cells (**Extended Data Fig. 5**). Furthermore, 3 km or higher resolution simulations produce intensive diabatic heating and $A_e$ to $K_e$ conversion by baroclinic conversion process [$\omega'\partial\Phi'/\partial p$] through more warm air rising in the eyewall from the lower troposphere to the upper troposphere[30,40], while less diabatic heating is released around the rainband region (**Supplementary Fig. 7**). This is supported by scale analyses of G($A_e$), $A_e$, C($A_e$, $K_e$) and $K_e$ in **Supplementary Text 1(b)** with two-way nesting downscalings at different resolutions. The statistics of asymmetry of potential vorticities[41] (see **Methods** and **Supplementary Text 2**) show that in the base 27 km simulations, only on the first 2 days do TFBs have a little stronger asymmetric structure than STEs but no difference is found for other times. In the 3 km downscaled simulations, the TFBs's asymmetry is always larger than the STEs's but much smaller than the difference between the 3 km downscaled simulation and base 27 km simulation (**Extended Data Fig. 6a**). While the 3 km or higher resolution simulation exhibits asymmetric fine-scale cloud structures (**Extended Data Fig. 5, Supplementary Figs. 3, 5-6, Extended Data Fig. 6a**), such asymmetric cloud structures produce asymmetric cyclonic flows (**Extended Data Fig. 6b,c**).

**Feedbacks of TC cumulonimbus clouds to environment controlling the turning of a TFB**

To elucidate how asymmetric momentum anomalies induced by fine-scale clouds correct the TFB's tracks, we first examine the process through which fine-scale cumulonimbus clouds trigger changes in TFB's track using TC potential vorticity (PV) analysis (see **Supplementary Text 2**). Since a TC tends to follow the maximum local PV tendency to move, we use the PV tendency gradient at wavenumber-1 (PVTG$_1$) to represent the TC movement vector[42-44] and analyze its budget as the contributions of diabatic heating from local clouds and horizontal advection in driving the TC motion[45]. We use Typhoons Kammuri and Krosa as examples, representing those with sudden changes in moving direction, to show how asymmetric distribution cumulonimbus clouds (**Extended Data Fig. 6b, c**) initiate the track change of TFBs. We compare the decomposition of the PVTG$_1$ vector in the 3 km downscaled simulation with the 9 km resolution counterpart to elucidate the triggering process (**Extended Data Fig. 7** and **Supplementary Fig. 9**) (see **Methods & Supplementary Text 3**). We find that it is the change of local horizontal advection that largely changes PVTG$_1$ and disturbs the TC-moving direction, which is governed by asymmetric momentum anomalies produced by the fine-scale cumulonimbus clouds. However, without feedbacks to the environment, such local cloud forcings cannot be transferred to the large-scale environment to ultimately change the TC track.

To clarify how the fine scale convection and environment mutually affect each other to determine the TC turning, we first show the increased accuracy in simulating Typhoon Kammuri's track at 3-km resolution nesting simulations as the feedback strength increases from null to full as a set of sensitivity experiments (**Fig. 4A**). This demonstrates the controlling role of feedbacks from the cumulonimbus clouds to the environment for the change of TC track. Then we examine the time evolution of VTFs and steering flows, the configuration of $\beta$-gyres and large-scale background, as well as the time series of decompositions of TC PVTG$_1$ with different feedback strengths (**Supplementary Fig. 9**). The analyses start from the early pre-turning stage of the TC, at which the TC's cumulonimbus cloud cells distribute densely to its north and sparsely to its south (**Extended Data Fig. 6b**), producing extra latitudinally-asymmetric flows according to the latitudinal variation of the Coriolis effect (see more detailed analyses in **Methods**). Such latitudinal variation of the Coriolis effect perturbation (stronger in the north and weaker in the south in this case) disturbs the TC $\beta$-gyres that tend to reorient cyclonically. These results clearly show that under relatively-weak environmental steering, such disturbances of the TC $\beta$-gyres, that link the TC clouds and environment[46-48], influence the environment which eventually establishes a new TC steering that controls the TC track. This mechanism can be shown in the change of typhoon tracks estimated by planetary vorticity advection (*pvAE*) as the strength of feedbacks of cumulonimbus clouds to the environment changes from null to full (**Fig.**

**4B**). A *pvAE* TC position is a pair of weighted mean longitude and latitude by planetary vorticity advection (*pvA*) in the TC-domain, which reflects the role of asymmetric *pvA* on TC track. The nearly-identical behaviors of the *pvAE*-track and the model TC track (compare **Fig. 4B** to **Fig. 4A**) clearly show that it is the asymmetric *pvA* produced by fine-scale clouds that makes the TC's turning.

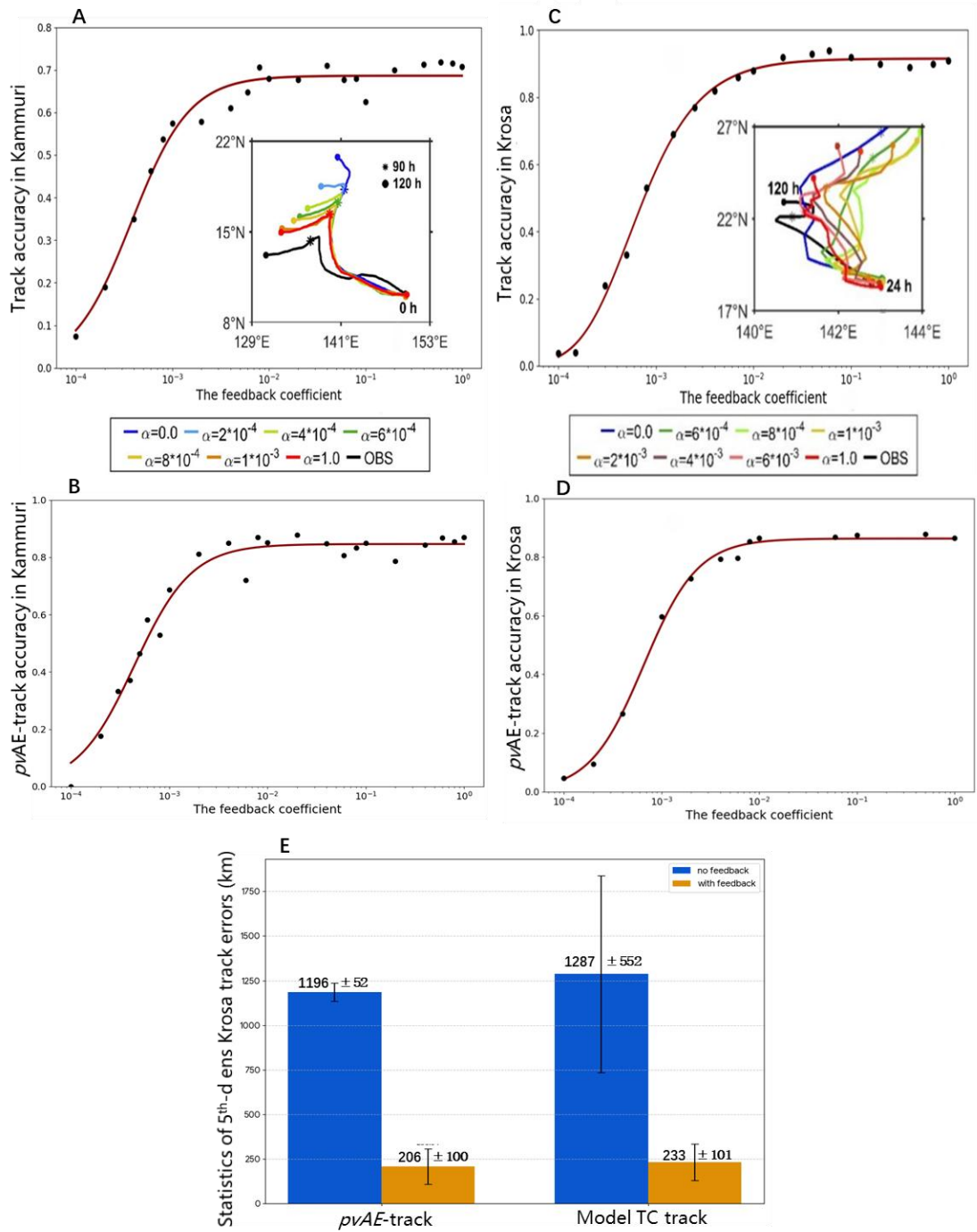

**Fig. 4 | Feedbacks of TC cumulonimbus clouds to environment controlling the turning of TFBs. A,** Variation of Kammuri's track accuracy with feedback strengths of fine-scale TC clouds to the environment. Track accuracy is

defined as $1-|\varepsilon_\alpha/\varepsilon_0|$ where $\varepsilon_\alpha/\varepsilon_0$ is the ratio of mean position errors of $96^{th}$-$120^{th}$ hours at feedback strength α over the one with no feedback (α=0) in TC-downscaling simulations. The small box shows 7 samples of simulated tracks with 7 different α values (see notes in the underneath box). **B,** The same as A, but for the track positions estimated by planetary vorticity advection (denoted as *pv*AE-track accuracy). A *pv*AE TC position is a pair of weighted mean longitude and latitude in the TC domain by instantaneous absolute *pvA* values (see **Methods**). **C**, The same as **A** but for Krosa. **D**, The same as **B** but for Krosa. E, The statistics of track accuracy of 21-member ensemble simulations of Krosa in 3km resolution cloud resolving simulations with feedbacks (brown) and 27km resolution base simulations (blue).

A similar story holds for Typhoon Krosa's turning in which the TC starts from asymmetrically-distributive clouds that are denser in the south and sparser in the north (**Extended Data Fig. 6c**), and it first turns from northwest to nearly due east, and then turns back to northwest (see **Supplementary Text 4**). While sensitivity experiments show that the feedback of cumulonimbus clouds to the environment controls the TC's turning (**Fig. 4C**), again, it is through the asymmetric *pvA* produced by the fine-scale clouds to carry out such turning (**Fig. 4D**). Both the model track and *pvAE*-track of TC show a nonlinear behavior with the strength of feedbacks of fine-scale clouds to the environment (**Fig. 4A-D**). We may comprehend this phenomenon as nonlinearity of the *pvA*'s response to fine-scale information in a very high frequency manner of information transfer between fine-scales and the background (every 30 second in this case). This is an interesting phenomenon that is definitely worth to be further studied in the future. To check robustness of the conclusion that it is the asymmetric *pvA* produced by the fine-scale clouds that makes TFBs' abnormal turning, we perform a set of multi-member ensemble experiments for Krosa which has a zigzag complex track. The 21-member ensemble of Krosa in 27km resolution base simulation and 27-9-3 km resolution full feedback downscaled simulation further confirm this mechanism **(Fig. 4E)** (see **Methods** for more detailed description and analyses**).** The processes that describe the turning mechanism can be visually shown in the animation of the evolution of Typhoons Kurmari's and Krosa's *β*-gyres and VTFs (**Supplementary Movies 1, 2, 3 & 4**).

**Summary and Discussions**

Aided by different TC-downscaled simulations with or without nesting that allows the downscaled information to influence the environmental flows, we find that resolving fine-scale clouds and their interactions with environment is essential for simulating the track change of TC track forecast busts sampled by typhoons (**Fig. 5A).** The cumulonimbus clouds organized by environment promote conversion of eddy available potential energy $A_e$ to eddy kinetic energy $K_e$ which sustains the evolution of the western ridge of the West Pacific subtropical high that steers the track. For simulations with broader parameterized clouds, diabatic heating overestimates the $A_e$ to $K_e$ conversion that distorts the western ridge of the West Pacific subtropical high, resulting in a different mechanism of the TC track change. The general physics of fine-scale TC clouds determining the correct track of busts by interacting with environment consist of three consecutive processes depicted in **Fig. 5B**. First, fine-scale convection produces asymmetric distribution cumulonimbus clouds that generate asymmetric momentum anomalies, while cloud diabatic heating leads to the conversion from eddy available potential energy $A_e$ to eddy kinetic energy $K_e$. Second, through the latitudinal dependency of the Coriolis effect, the cloud-induced asymmetric momentum anomalies reorient the $\beta$-gyres that are pre-conditioned by the large-scale environment, which can momentarily change the TC moving through the ventilation flows of the $\beta$-gyres. Third, through vorticity advection, the environment responds rapidly to the changes of $\beta$-gyres and develops new TC steering flows that in return contribute to the change of the $\beta$-gyres and controls the TC track.

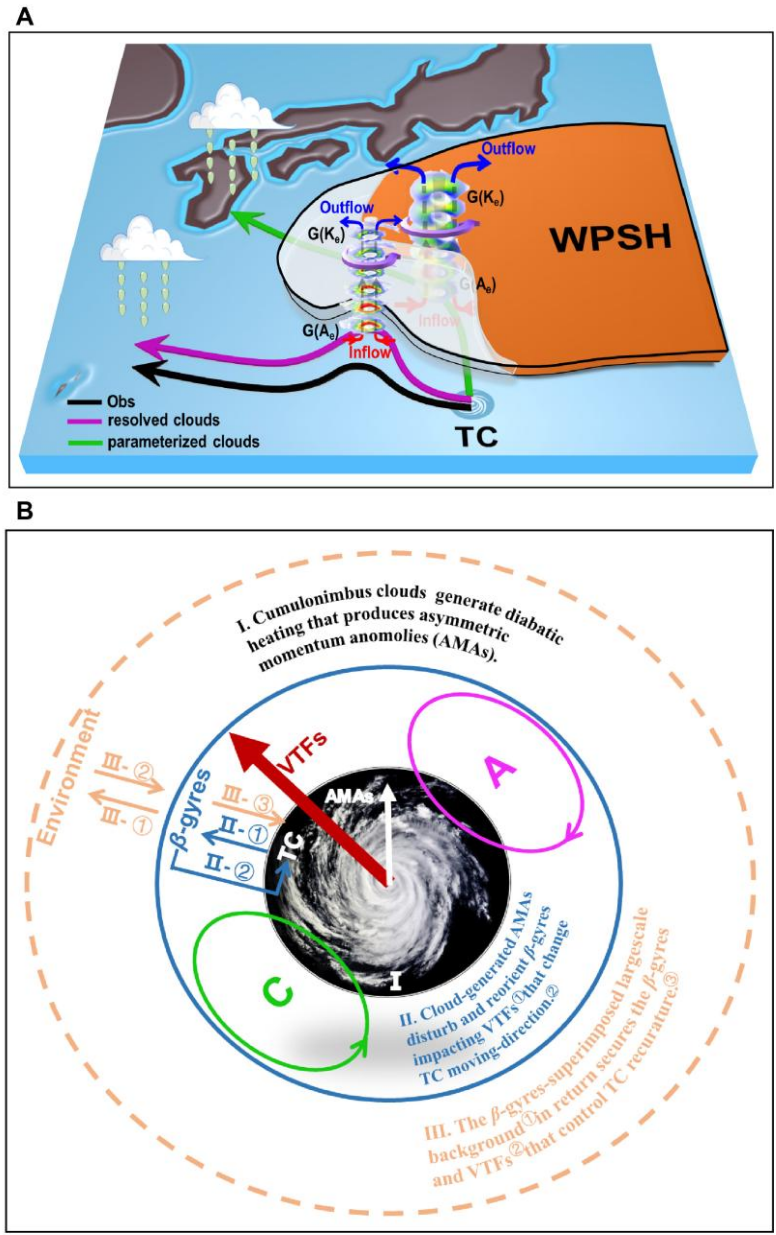

**Fig. 5 | Schematic illustration of physics in feedbacks of fine-scale clouds to environment regulating typhoon tracks. A,** Exhibition of clouds-environment interactions determining TFBs tracks in the West Pacific. Under a weak WPSH environment, parameterized clouds overestimate spatial coverage but underestimate central intensity, generating wide eddy available potential energy production [G($A_e$)] with weak central intensity. The resulted baroclinic energy conversion makes excessive eddy kinetic energy production [G($K_e$)], eroding the west ridge structure (white transparent shadow) of WPSH and leading to its unrealistic eastward retreatment, resulting in incorrect environmental flows that mis-steer TC tracks (green). Instead, resolved cumulonimbus clouds and their feedbacks to environment well organize the TC spiral cloud system with more realistic central intensity. The G($A_e$) and G($K_e$) by such fine-scale clouds have a limited impairment on WPSH, generating correct environmental flows that steer TC motion (purple) mostly as observed (black). **B,** Three consecutive processes linking fine-scale clouds with environment together to characterize their interactions: I, Cumulonimbus clouds generate diabatic heating that produces asymmetric momentum anomalies (AMAs; thick-white arrow), corresponding energy conversion of eddy

available potential energy ($A_e$) to eddy kinetic energy ($K_e$); II, Cloud-generated AMAs disturb and reorient $β$-gyres (purple and green circles with arrows) impacting ventilation flows (VTFs) (thick-red arrow) that change TC moving-direction momentarily; III, the $β$-gyres-superimposed largescale background in return secures the $β$-gyres and VTFs and controls TC track. The black, blue and orange circles denote the spatial scale of TC, $β$-gyres and the environment, respectively.

Although based on typhoons, our new finding that positive feedbacks between fine-scale clouds and environment could dramatically change the TC track underscores such feedbacks as a source of TC track predictability, which may also have a substantial impact on other TC-relevant fields (e.g. intensity). Based on such understanding of TC track predictability source, continuous improvement on TC track prediction is possible as TC finer-scale characteristics such as inner cloud convective updrafts and their interactions with the environment including underlying ocean conditions are better resolved as the coupled model has improved its behaviors on multi-scale interactions. While the mechanism of fine-scale cloud structures impacting TC intensity in a warmer climate has been explored[49,50], our new understanding of TC track predictability suggests that it is crucial that any assessment of the impact of future warmer climate on TC track must be carried out in the presence of feedbacks between fine-scale clouds and environment. Given the linkage role of TCs from weather to climate[51,52], our unique understanding on TC track predictability also serves as an important step in quantifying seamless weather-climate predictability.

**REFERENCE AND NOTE**

**Acknowledgments**

This research is supported by the National Natural Science Foundation of China (Grant Nos. 42361164616, 42022041) and Science and Technology Innovation Project of Laoshan Laboratory (grant nos. LSKJ202300400-03, LSKJ202202200-04) and Shandong Province "Taishan" Scientist Program (ts201712017).

**Author contributions**

S.Z. led and organized the research project, and S.Z., Y.G. and W.L. designed the study. H.Z. conducted experiments and H.Z. and Y.C. processed the data. H.Z., S.Z. and Y.C. performed the analyses. S.Z., Y.G., W.L. and H.Z. wrote the initial manuscript. C.W., B.W. and R.L. provided significant discussions to improve the initial manuscript. All other coauthors contributed to interpreting the results and improving the paper.

## Methods

**Unique nesting strategy to control cloud's feedback strength to environment.**

A regional coupled prediction system for the Asia-Pacific (AP-RCP) (16°S-63°N, 38°E-178°E) area has been established to make quasi-operational extended-range (≤35 days) predictions starting on January 1, 2018. The AP-RCP system configuration includes the Weather Research and Forecasting (WRF, v3.7.1) model at 27 km horizontal resolution and Regional Ocean Model System (ROMS, v3.7) at 9 km horizontal resolution. The coupled model covers a large domain of the Asia-Pacific, which extends far enough to simulate the generation and development of TCs in the West Pacific. WRF is divided into 28 vertical levels from the surface to the model top at 50 hPa. The primary physics schemed selected in WRF includes (i) the Kain-Fritsch convection parameterization scheme[54]; (ii) the Rapid Radiative Transfer Model for GCMs (RRTMG) longwave and shortwave radiations[55]; (iii) the Yonsei University (YSU) boundary layer scheme[56]; (iv) the WRF single-moment 3-class (WSM3) microphysics scheme[57]. Detailed description of the ROMS can be found in reference[28]. The initial and boundary conditions of the AP-RCP system are obtained from the Climate Forecast System Version 2 (CFSv2) product as the background accompanied with the standard 3-dimensional variational data assimilation in the atmosphere and a multiscale ocean data assimilation scheme for dynamically downscaled coupled data assimilation initialization[58].

While the coupled AP-RCP system is a good platform to study the many aspects of TC such as TC genesis predictability[58], we only take its atmospheric component to study the impact of fine-scale clouds on TC track predictability. Starting from the AP-RCP's atmosphere component with the TC-resolving horizontal resolution (27 km), we set multilayer nesting in the TC area (**fig. S2**) to examine the impact of the interaction between resolved-scales clouds and environmental flows on totally 8 TC track forecast busts (TFBs) with large track errors. The model domains are nested in 4 layers, with horizontal resolutions of 27 (D01), 9 (D02), 3 (D03) and 1 (D04) km and dimensions of 559×409, 219×219, 441×441 and 582×582, respectively. For the 3 km and higher resolution simulations, the cumulus convection parameterization is turned off. The TC-following method[59] is applied to the D02, D03 and D04, while the outermost D01 is fixed over the AP area. All these embedded sub-domains are centered at the TC core and cover the whole TC area. The initial and boundary conditions of the nested fine-scale simulations are obtained from AP-RCP. The role of fine cross-scale interactions between a TC and the environment in the regulation of TC track is investigated by multi-layer one-way or two-way nested TC-moving downscaling simulations. That is saying that the simulation results on higher-resolution (HR) nesting domain which contain smaller-scale information can impact back on the lower-resolution (LR) simulation by a nudging update formula:

$$x_{lr}^{u} = (1-\alpha)x_{lr} + \alpha \bar{x}_{hr}. \quad (1)$$

Here, $\bar{x}_{hr}$ represents the average value of HR simulation results on a LR grid-box, while $x_{lr}$ and $x_{lr}^{u}$ represent the LR model simulation and updated results respectively, and α is the feedback coefficient of HR to the LR simulation with 0 (1) as one-way (full two-way) nesting without feedback (with full feedback), and a value between 0 and 1 controls the strength of feedback.

**Detection of TC track sensitivity to environmental conditions.**

The movement of a TC is largely governed by its surrounding deep-layer-mean environmental flow, i.e. the steering flow. While the steering flow of a TC is typically calculated as the vertical mean of azimuthally averaged winds within the 5°~7° radial band from the TC center between 850 and 200 hPa[60], some studies found that the averaged wind within a radius of 5° from the TC center is more consistent with the TC motion. Such uncertainty about the horizontal domain over which the flow should be integrated makes it difficult to precise represent the characteristics of the large-scale environment. The Western Pacific Subtropical High (WPSH) is a prime large-scale synoptic pattern that affects western North Pacific TC (i.e. typhoons) activities[61] and therefore it robustly represents the large-scale environment of typhoons. The extension and withdrawal of WPSH have strong relationship with TC movement[62]. In order to efficiently characterize variations of the WPSH, following the previous studies[61], here we calculate the indices for WPSH strength and area as follows, which have been widely used by the National Climate Center in China. The WPSH strength index is the sum of 500 hPa Geopotential Height (GHT) values greater than 5880 gpm substracted by 5870 gpm within the range of 40°×40°centered at each TC, while the WPSH area index is the total area of grid boxes for which the 500 hPa GHT value is greater than 5880 gpm within the range of 40°×40° centered at each TC. Combining these two indices with traditional steering flow, we can have a better representation of large-scale environmental steering flow for a TC and more readily to understand the relationship between large-scale environment and TC motion (**Fig. 2** and **Extended Data Fig. 3**).

To detect the impact of the large-scale environmental flow on TC motion, a filtering method[63] is applied to separating the original model state field $F$ including the surface pressure, $u$ and $v$ components of wind, temperature and GHT into the basic large-scale general feature $F_B$ and the departure from $F_B$, called $F_D$ as:

$$F = F_B + F_D. \tag{2}$$

$F_B$ is obtained by a local three-point smoothing operator. Then, the initial strength of a large-scale system can be changed as $F_{new}=F+\gamma \times F_D$ so that the strength of environmental flow can be controlled by the coefficient $\gamma$. Here we set the coefficient $\gamma$ as 0, .5 and 2 to represent three strengths of WPSH. While the reference ($\gamma$=0) 500 hPa winds and GHTs at the onset time of Kammuri in the base 27km-res simulation is shown in **Extended Data Figs. 4a**, the other two cases of WPSH strength are shown in **Extended Data Figs. 4b** as green ($\gamma$=.5) and red ($\gamma$=2). Under the circumstances of 3 different strength environmental flows, the model tracks of Kammuri in base 27km-res (solid) and 3km-res downscaled (dotted) simulations are shown in **Extended Data Figs. 4c** as purple, green and red color respectively. At the reference strength ($\gamma$=0), the base 27km-res simulation differs largely from the 3km-res downscaled simulation and the observation. As the strength of WPSH environmental flow is enhanced, the difference among the base 27km-res simulation, 3km-res downscaled simulation and observation becomes small. This numerical experiment results support the comprehension that under relatively weak environment, the interaction between fine-scale clouds and environment plays more important role in determining TC motion.

**Wavenumber spectra analysis.**

The horizontal wavenumber spectra of radar reflectivity and satellite-observed cloud optical thickness (COT) in **Fig. 3** are computed as[35]:

$$E(k_h) = \frac{1}{2} \cdot \sum_{k_h-\Delta k}^{k_h+\Delta k} \frac{[\hat{q}(\boldsymbol{k})\hat{q}(\boldsymbol{k})^*]}{\Delta k}, \tag{3}$$

where $\hat{q}(\boldsymbol{k})$ is the Fourier coefficient of targeted variable $q$, $\boldsymbol{k}=(k_x, k_y)$ is the horizontal wave number vector, * is the complex conjugate, and $k_h = |\mathbf{k}| = \sqrt{k_x^2 + k_y^2}$ is the total horizontal wavenumber. Given the two-dimensional energy spectra, one-dimensional $k_h$ spectra are constructed by angular averaging over wavenumber bands $k_h - \Delta k \leq |\mathbf{k}| \leq k_h + \Delta k$ on the $k_x - k_y$ plane, where $\Delta k = 2\pi/(\Delta s \times N)$ and $N = min(N_i, N_j)$, and $N_i$ and $N_j$ are the number of zonal and meridional grid points, respectively, $\Delta s$ is the horizontal grid spacing.

**Generation of eddy available potential energy $A_e$ and eddy kinetic energy $K_e$.**

The generation of $A_e$ [i.e. G($A_e$)] and eddy baroclinic energy conversion from $A_e$ to $K_e$ [i.e. C($A_e$, $K_e$)] in **Extended Data Figs. 5** are computed as:

$$G(A_e) = \frac{1}{g} \int_{P_s}^{P_u} N' \dot{Q}' dp, \tag{4}$$

$$C(A_e, K_e) = -\frac{1}{g} \int_{P_s}^{P_u} \frac{\partial \omega \phi'}{\partial p} dp, \tag{5}$$

where $g$ is the acceleration of gravity, $N$ is efficiency factor, $\dot{Q}$ is diabatic heating rate, $p$ is pressure, $\omega$ is vertical motion in isobaric coordinates, $\phi$ is geopotential height, $p_s$ and $p_u$ are the surface and upper pressure level, and ()' represents the perturbation referred to the area-mean. The area for computing perturbation is defined as a 6°×6° square centered at the TC. The terms in the equations are vertically integrated from 1000 hPa to 100 hPa. The detailed equations of both $A_e$ and $K_e$ are used to analyze the eddy energy budget in **Supplementary Text 1(b)**.

**The asymmetry of potential vorticity distribution.**

An index that measures the asymmetry of potential vorticity is defined as[41,48]:

$$\gamma(r,z,t) = \frac{\int_0^{2\pi} P'(r,\lambda,z,t)^2 d\lambda/2\pi}{\overline{P}(r,z,t)^2 + \int_0^{2\pi} P'(r,\lambda,z,t)^2 d\lambda/2\pi}, \tag{6}$$

where $P$ denotes potential vorticity, $r, \lambda, z, t$ are the radius, azimuth, height and time, respectively. An overbar and prime represents the azimuthal mean and the departure of local value from the mean. The index $\gamma$ is calculated as the average of vertically 850-200 hPa and horizontally within a 300 km radius.

**Potential vorticity tendency gradient (PVTG) analysis for TC motion.**

The wavenumber-1 (WN1) of local PVTG, the maximum asymmetric component of PVTG, called $PVTG_1$, can be used to define TC moving direction and speed[45]. By processing potential vorticity tendency equation under azimuthal WN1 analysis technique, the $PVTG_1$ vector can be calculated as the sum of the contributions as:

$$(\nabla \frac{\partial P}{\partial t})_1 = \Lambda_1 [-\nabla(\vec{V} \cdot \nabla P) - \nabla(\omega \frac{\partial P}{\partial p}) - \nabla g \nabla_3 \cdot \left(-\frac{\dot{Q}}{C_p \pi} \vec{q} + \nabla \theta \times \vec{F}\right)], \quad (7)$$

where the first term at righthand side is the horizontal advection (HA), the second term is vertical advection (VA), the third term is diabatic heating (DH) and the forth term is friction (FR). In Eq. (7), $P$ is potential vorticity, the subscript "1" indicates the azimuthal WN1, $V$ is the horizontal air motion, $\omega$ is the vertical velocities, $p$ is the pressure, $g$ is the gravitational acceleration, $Q$ is the diabatic heating rate, $C_p$ is the specific heat of dry air at constant pressure, $q$ is the absolute vorticity, $\theta$ is potential temperature, $F$ is the friction. $\nabla$ and $\nabla_3$ denote the horizontal and three-dimensional gradient operators, respectively. All calculations are averaged within a radius of 500 km from the TC center and between 850 and 200 hPa. Ignoring friction and dissipation, we can quantitatively evaluate the contributions of nonlocal advection and local cloud diabatic heating in driving TC moving. Following the detailed diagnostic process given in **Methods** and **Supplementary Text 2,** the contribution of each factor, HA, VA and DH can be computed. The steering by environmental flow is included in HA, while VA and DH mainly depend on the dynamic and thermodynamic structure of the simulated TC.

**Cumulonimbus clouds triggering TC track changing in Typhoon Kammuri.**

To explore the role of fine-scale clouds in TC turning, we examine the simulated tracks of Typhoon Kammuri (2019) in different fine-scale downscaling schemes (**Extended Data Figs. 7a**). The Typhoon Kammuri has a sharp direction change that exceeds 90° in 12 hours. While a 3 km or finer resolution two-way nesting simulates the sharp northeast-to-southwest turning very well, the one-way nesting downscaling only simulates a small degree and much later turning with slower TC moving. This suggests that, without feeding fine-scale information back to large-scale background, local cumulonimbus clouds only play some triggering role for TC turning.

We first check the TC cloud structure simulated by 3 km resolution downscaling, which have a distinctive asymmetric distribution feature of cumulonimbus clouds that are dense in north and sparse in south at pre-turning time (**Extended Data Figs. 6b**). To understand the process of fine-scale clouds to trigger TC turning, we further examine the difference of behaviors of 3 km and 9 km resolution one-way nesting simulations. We first conduct analyses of TC potential vorticity. At the early period of pre-turning ($55^{th}$-$70^{th}$ hours), the 3 km resolution model TC first produces different local cloud $DH_1$, then changes $HA_1$ and departs from the 9 km resolution counterpart. But without sending fine-scale information back to mother-domain, it tends to merge back to the 9 km resolution track and the turning at $78^{th}$-$85^{th}$ hours does not occur.

Compared to the 9 km resolution counterpart, the 3 km resolution spiral cloud system (rainwater mixing ratio distributions) rotates cyclonically more rapidly (after the $50^{th}$ hour) and exhibits a finer scale and more asymmetric feature (**Extended Data Figs. 7c**). For the $PVTG_1$ vector, we can compare the behavior of 3 km resolution downscaling simulation to 9 km counterpart, and examine the contribution directly from asymmetrically-distributive cloud diabatic heating ($DH_1$) and the contribution from the induced PV advection ($HA_1$) (**Extended Data Figs. 7d**). From 3 km resolution to 9 km resolution downscalings, as the consequence of spiral cloud structures shown in **Extended Data Figs. 7c**, the direct contribution of $DH_1$ for TC $PVTG_1$ is relatively small (yellow). It is the change of $HA_1$ associated with the asymmetric momentum anomalies produced by fine-scale cumulonimbus clouds that largely makes the change of $PVTG_1$ (brown to blue) and disturbs TC-moving direction at early pre-turning as shown in **Extended Data Figs. 7b.**

**Abnormal turning process of Typhoon Kammuri.**

At the early pre-turning stage, the Kammuri's cumulonimbus clouds distribute densely to its north and sparsely to its south, producing extra asymmetric flows (**Extended Data Figs. 6b**) according to the latitudinal dependency of the Coriolis effect. Such perturbation of Coriolis effect from asymmetric distribution clouds, stronger in the north and weaker in the south, disturbs the TC *β*-gyres that tend to be reoriented cyclonically. Under relatively-weak environmental steering, once such disturbances influence the environment, the TC *β*-gyres that link the TC clouds and environment start to rotate cyclonically (dotted in **fig. 16a**) and the TC goes in its pre-turning phase. The stronger the feedback, the sooner the reorientation of the *β*-gyres begins, highlighting the governing role of asymmetric local conditions in influencing the direction of VTF after the fine-scale TC-convection exerts its impact on the environment. During this period, asymmetric momentum anomalies produced by fine-scale TC clouds propagate to the large-scale background, favoring TC-turning as the spreading effects of fine-scale convections accumulate. The VTF magnitude increases but its increasing rate relative to the increasing rate of steering (TC steering flow subtracted by VTF) is smaller (dashed-dotted) whereas the direction of TC steering flow is not yet significantly changed by the feedback (solid).

As the environment continues to be influenced by the fine-scale convection, the TC *β*-gyres are strengthened and the increasing rate of VTF magnitude exceeds that of the steering, and the TC enters its in-turning phase (from the $78^{th}$-hour on) (dashed-dotted in **fig. S16a**). While the TC *β*-gyres continue to rotate cyclonically and be strengthened after the $86^{th}$-hour, the positive feedback between fine-scale convection and environment amplifies, and the TC-steering rapidly changes its direction (**solid**) and the TC dramatically turns. In just a few hours, as the TC *β*-gyres and the western ridge structure of WPSH are tightly connected (**fig. S16b**), the TC rapidly completes its turning at the $92^{nd}$-hour. The co-evolution of TC *β*-gyres (color-dotted) and the WPSH environment (color-solid) clearly shows the process of fine-scale TC clouds feeding back to the environment in determining the TC turning. In this process, the leading role of TC *β*-gyres can be further confirmed by the lag behavior of the WPSH in the co-evolution of TC β-gyres and the WPSH shown at different times and for different feedback strengths (**Extended Data Fig. 8**).

The interaction process of fine-scale TC convection and environment in TC turning can be further understood by analyzing the time evolution of $PVTG_1$ vectors that are decomposed as contributions of $HA_1$ and $DH_1$ (**fig. S16c**) in simulations with (upper dashed-box) or without (lower dashed-box) feedbacks. For the simulated turning with feedbacks, the $PVTG_1$ vectors (purple) better represent the TC movement than the traditional steering flows (black) which are consistent with $HA1$ (blue). In the two-way nesting case, the feedback of asymmetric diabatic heating produced by fine-scale convection (yellow) first affects the $HA_1$'s VTF part (green) and its impact is then transferred to $PVTG_1$ (purple). With feedbacks, the *β*-gyres' change is linked with the environment through advection. The TC-convection favorable environment in turn produces asymmetric distribution fine-scale convection cells ($DH_1$ vector) as well as the configuration of the *β*-gyres (VTF vector) and TC background (steering vector). Hence $DH_1$, $HA_1$ and $PVTG_1$ tend have similar directions (see $90^{th}$-$93^{th}$ hours in the upper dashed-box), and the TC completes its turning. In the one-way nesting case, since the information transfer mechanism is absent, the fine-scale convections (brown) do not affect the $PVTG_1$ vectors (magenta arrow), and $DH_1$, $HA_1$, VTF do not link to $PVTG_1$.

The co-evolution of TC spiral clouds and induced asymmetric momentum anomalies in the 3 km resolution simulation compared to the 9 km resolution case can be visually viewed in **Movie S1**. The evolution of *β*-gyres in 3 km resolution one-way and two-way downscaled simulations can be visually viewed in **Movie S2**. Both movies strongly support the analyses above.

**The typhoon track estimated by planetary vorticity advection.**

Inspired by previous study of planetary vorticity advection (*pvA*) on TC track[48], here we define a *pvAE*-track to measure the role of asymmetric distribution *pvA* on typhoon tracks. Each TC position in a *pvAE*-track is a pair of weighted mean longitude and latitude by *pvA* as

$$\theta_{pvAE} = \frac{\sum |pvA_i| \times \theta_i}{\sum |pvA_i|}. \tag{8}$$

Here $\theta$ denotes longitude $\lambda$ (or latitude $\varphi$), $i$ is the grid-box index, $|pvA_i|$ is the absolute value of area mean of *pvA* at the i$^{th}$ grid-box, and $\sum$ represents the sum in the whole TC-domain (the innermost 3 km resolution domain in this case). Then, a pair of ($\lambda_{pvAE}, \varphi_{pvAE}$) represents a TC position that is estimated by the asymmetrically-distributive *pvA*.

The variation of *pvAE*-track [i.e. the TC track estimated by Eq. (8)] with the feedback strength is nearly identical to the variation of traditional model TC track defined by minimum surface pressure etc. multiple factors[64] in both Kammuri and Krosa cases and both show a nonlinear behavior (compare **Fig. 4C-D** to **Fig. 4A-B**). We may comprehend this phenomenon as nonlinearity of the response of planetary vorticity advection to the feedback of fine-scales, resulting from very high frequent information transfer (every 30 second in this case) between fine-scales and the background. This is an interesting phenomenon that is definitely worth to be further studied in the future. To check robustness of the mechanism that the asymmetric *pvA* produced by the fine-scale clouds makes TFBs' abnormal turning, we perform a set of multi-member ensemble experiments for Krosa which has a zigzag complex track, based on the GEFS/NCEP (Global Ensemble Forecast System/National Centers for Environmental Prediction). This ensemble experiment generates 21 members with perturbations in the initial conditions for the global circulations, which are used to provide the initial and boundary conditions for our downscaling simulations. Following the methodology of 27km resolution base simulation and 27-9-3km resolution full feedback downscaled simulation, two groups of 21-member ensemble simulations are conducted for Krosa. The 27km resolution base simulation and 27-9-3km resolution full-feedback ensembles (**Fig. 4E**) give completely consistent results with the above sensitivity experiments. From **Fig. 4**, we also learned that the *pvAE*-track is closer to the observation than the modeled TC track estimated by the traditional multi-factor way in all sensitivity and ensemble experiments and the standard deviation of *pvAE*-track is smaller, especially in in the base simulation (see error bars in **Fig. 4E**). This is because of the dominant role of planetary vorticity advection in TC track, while the actual model

TC track reflects a TC's complex consequences on multiple aspects. In the base simulation ensemble, the effect of identical convection parameterization is much dominant over the perturbations of initial and boundary conditions in *pvA*.

**Data Availability Statement**

The ERA5 reanalysis data is downloaded from https://cds.climate. copernicus.eu/cdsapp#!/search?text=ERA5. The iBTrACS best track data can be accessed at https://climatedataguide.ucar.edu/climate-data/ibtracs-tropical-cyclone-best-track-data. Himawari-8 data of TC structure can be accessed at https://www.eorc.jaxa.jp/ptree/index.html. All simualted model data are available from the corresponding author on request.

**Codes Availability**

All model and analysis codes are available from the corresponding author on request.

**Methods References**

**List of 8 Extended Data Figures:**

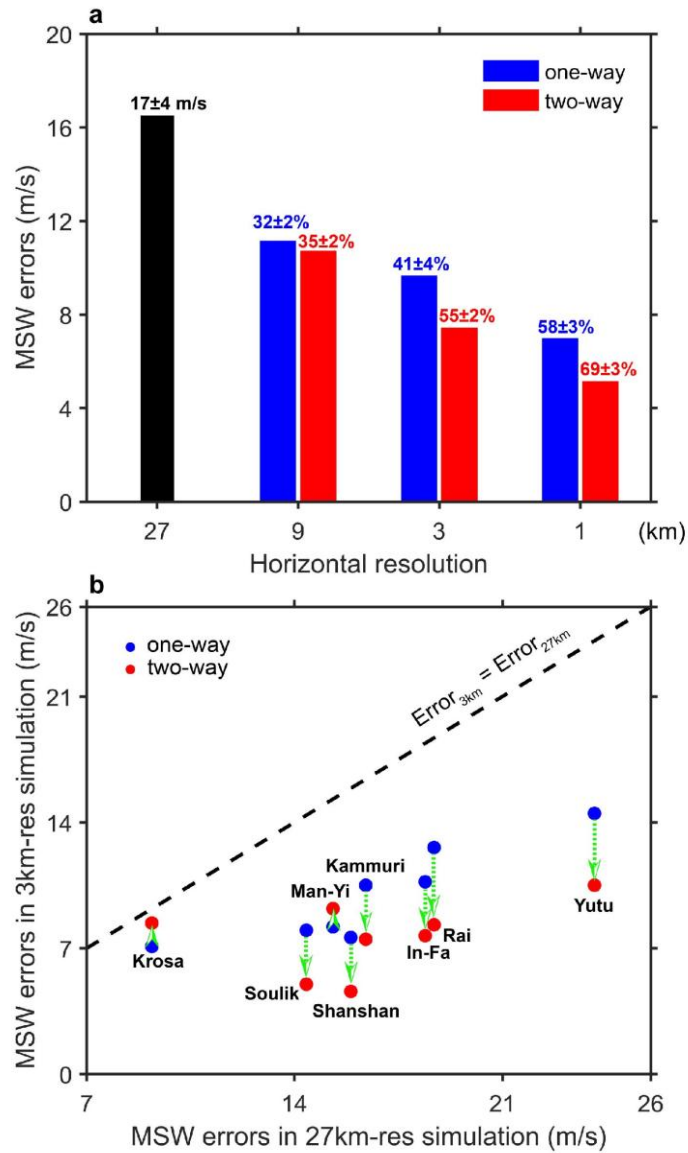

**Extended Data Figure 1 | Model evidence of fine-scale TC-environment interactions improving TC intensity prediction. a, b,** Same as **Figs. 1b**, **c** but for the mean absolute errors of the 5th-day predicted maximum sustained wind (MSW) (m/s). The black-dashed line indicates the bound at which the MSW errors by 3km-res downscalings and base 27km-res simulations are equal. The downward (upward) arrows indicate the reduction (increase) of MSW error.

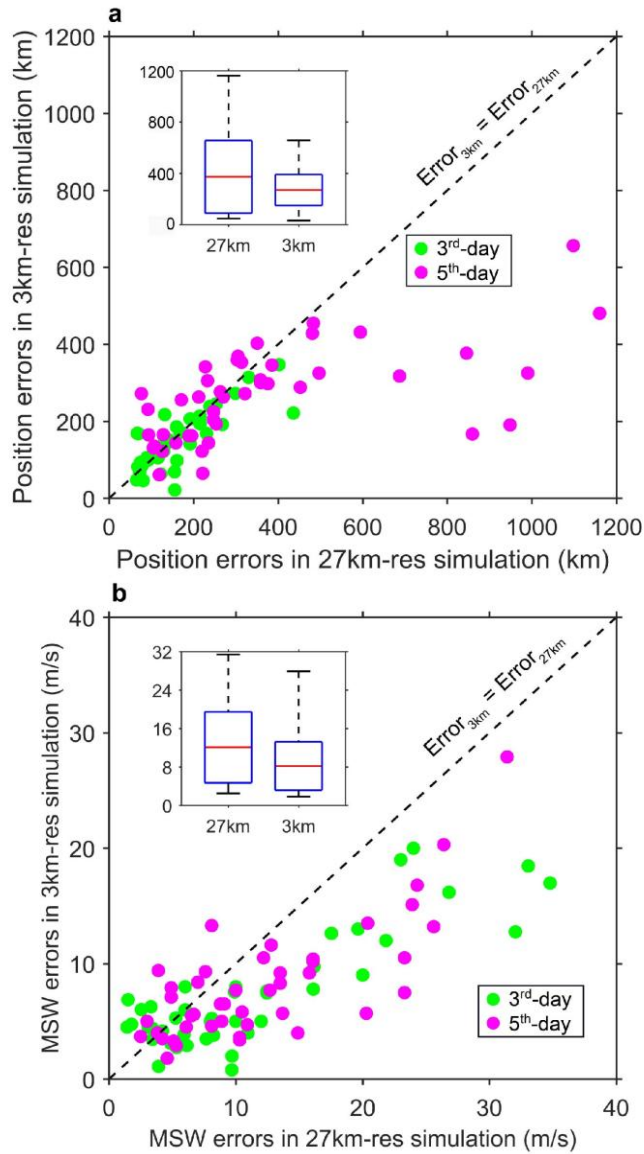

**Extended Data Figure 2 | The improved TC predictability by resolving fine-scale TC-environment interactions in 46 typhoons. a, b,** Scatterplots of position (panel **a**) and MSW (panel **b**) errors for the 3rd-day (green) and 5th-day (pink) predictions in base 27km-res simulations (x-axis) vs. 3km-res downscalings (y-axis). Boxplots in each panel are the corresponding uncertainty estimation of the 5th-day predictions in 43 cases, in which the central red lines denote the mean and blue boxes show the standard deviation, and lower and upper whiskers denote minimum and maximum values.

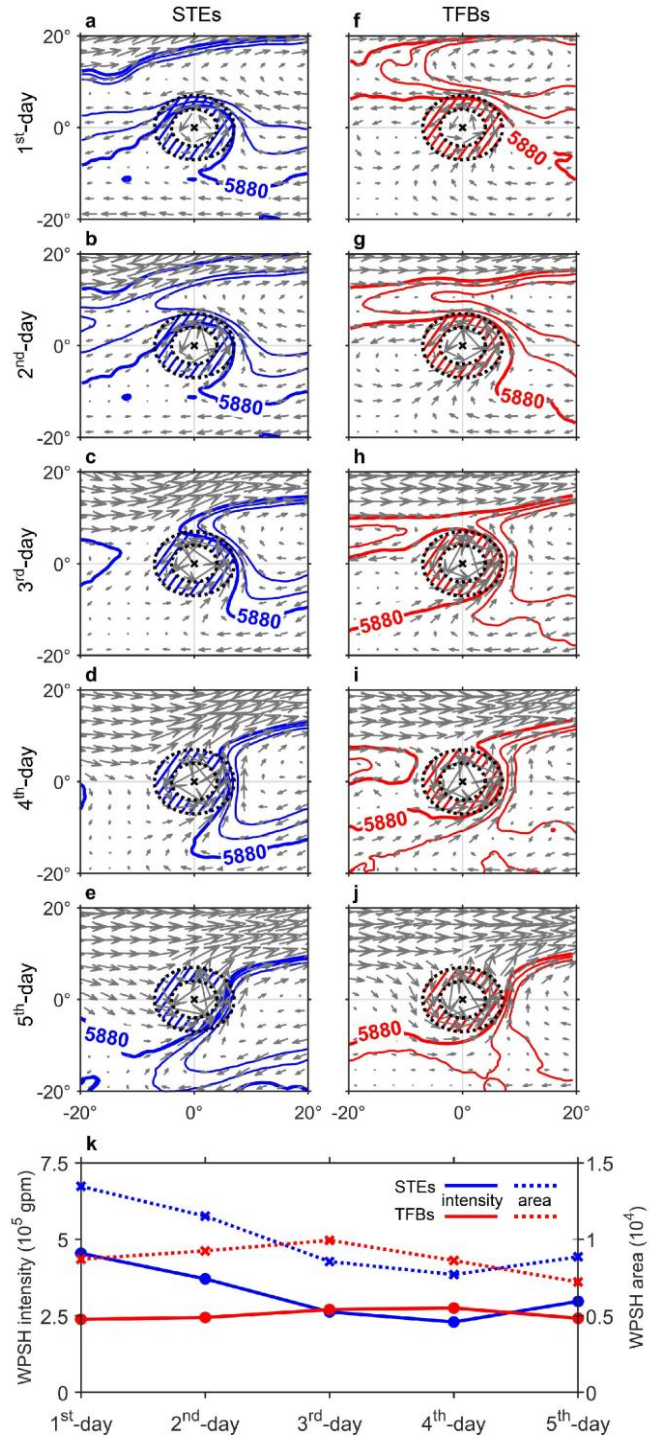

**Extended Data Figure 3 | The large-scale environment in TFBs vs. STEs. a-j**, The first 5-day 500 hPa winds and geopotential height composited with respect to TC centers for 8 STEs (left: **a-e**) and 8 TFBs (right: **f-j**) in base 27km-res simulations. Contour intervals are 7 gpm and contours less than 5880 gpm are not shown. The domains are 40° latitudes×40° longitudes centered at TC eye. Hatched areas represent calculation region of steering flows in a radius range of 400-700 km from TC center. **k**, Timeseries of daily WPSH strengths (solid) and areas (dotted) computed

from 8 STEs (blue) and 8 TFBs (red).

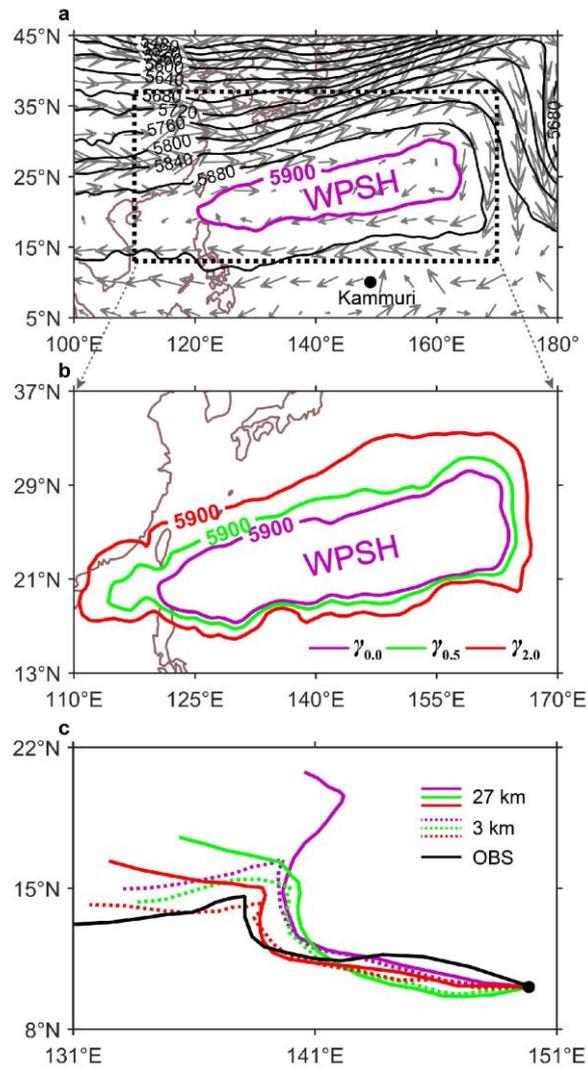

**Extended Data Figure 4 | The simulated tracks of Typhoon Kammuri by initializing model with different strength WPSH environments. a,** The 500 hPa winds and geopotential heights (GHTs) at the onset time of Kammuri (0000UTC 26 Nov. 2019) in the base 27km-res simulation. The WPSH strength is represented by the area circumscribed by the 5900 gpm GHT contour (purple). **b,** The sketch map of three WPSH strengths controlled by the disturbance coefficient $\gamma$ [$\gamma$=0.0 (original strength), $\gamma$=0.5 (moderate strength), and $\gamma$ =2.0 (high strength)] at 0000UTC 26 Nov. 2019. **c,** The simulated tracks of Typhoon Kammuri in base 27km-res simulations with 3 strength WPSH environments shown in panel **b** (solid) and corresponding 3km-res two-way downscalings (dotted).

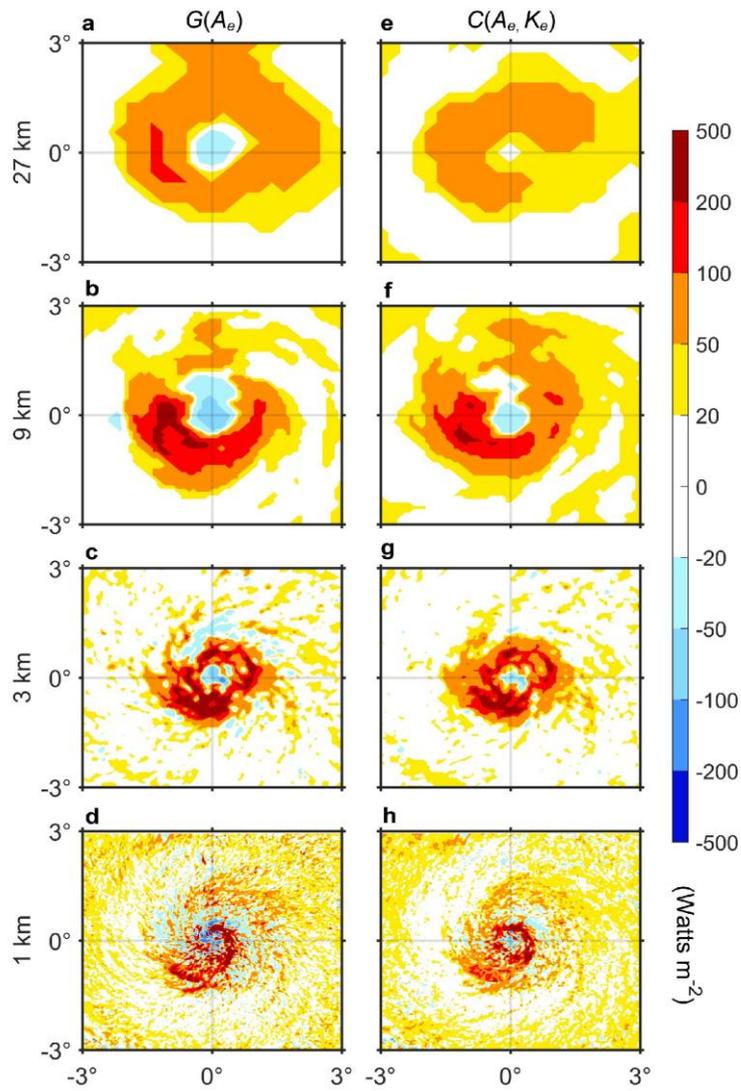

**Extended Data Figure 5 | 3km- or finer-res downscalings producing energy production consistent with satellite-observed cloud structures. a-h,** The composite analyses of 8 TFBs for diabatic heating representing eddy available potential energy generation G($A_e$) (left: **a-d**) and baroclinic conversion from eddy available potential energy to eddy kinetic energy C($A_e$, $K_e$) (right: **e-h**) representing major eddy kinetic energy generation G($K_e$), at the 96th-hour predictions in base 27km-res simulation, and 9km-, 3km- and 1km-res with feedback two-way downscalings.

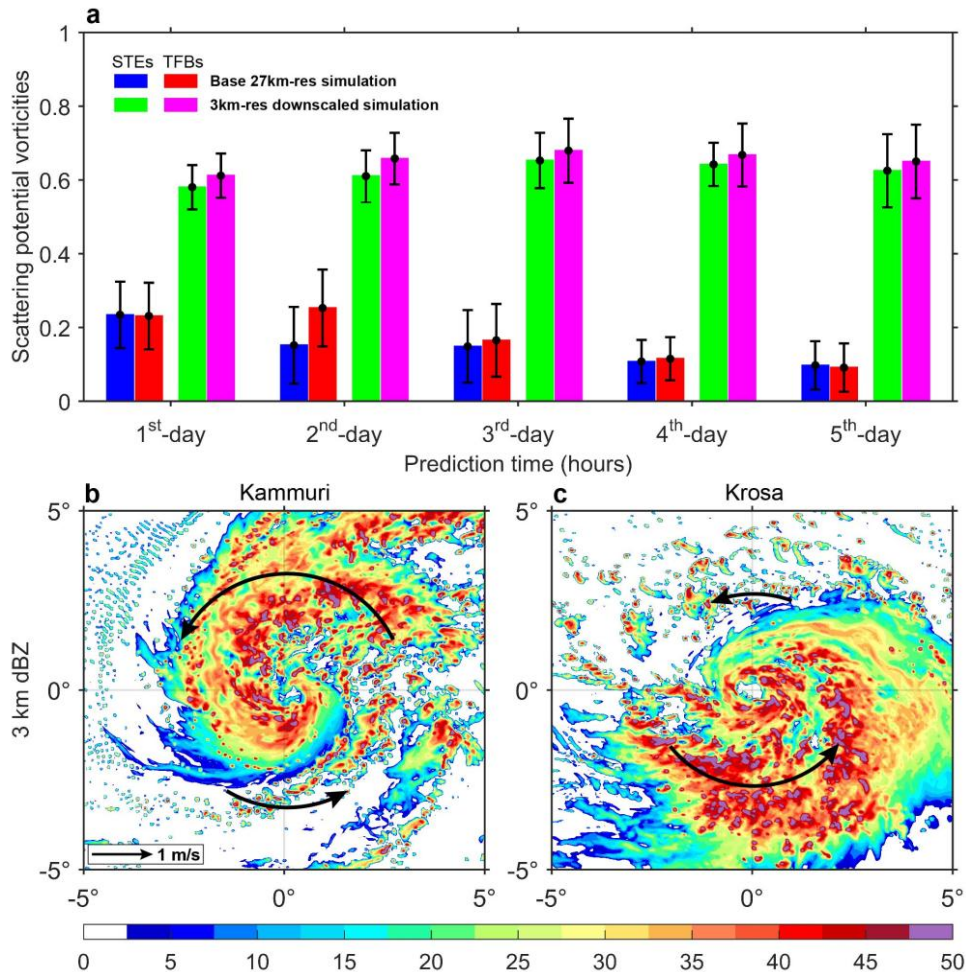

**Extended Data Figure 6 | Unevenly-scattering cumulonimbus clouds driving TC abnormal turning by reorienting TC *β*-gyres. a,** Grouped bar charts of 8-typhoon composite analyses of scattering potential vorticities for TFBs (red and purple) vs. STEs (blue and green) in base 27km-res simulations (blue and red), and 3km-res two-way downscaled simulations (green and purple) by the first 5-day predictions. **b, c,** The distributions of 3km-res downscaled simulation dBZ for Typhoons Kammuri (panel **b**) and Krosa (panel **c**) at pre-turning time (50th-hour for Kammuri and 34th-hour for Krosa). The black-curved arrows indicate latitudinally-asymmetric cyclonic flows induced by unevenly-scattering cumulonimbus cloud cells inside the TC, computed as the 850-200 hPa mean flow in the north (south) part of the TC. The vertical black dots and segments denote the mean and standard deviation in 8 cases.

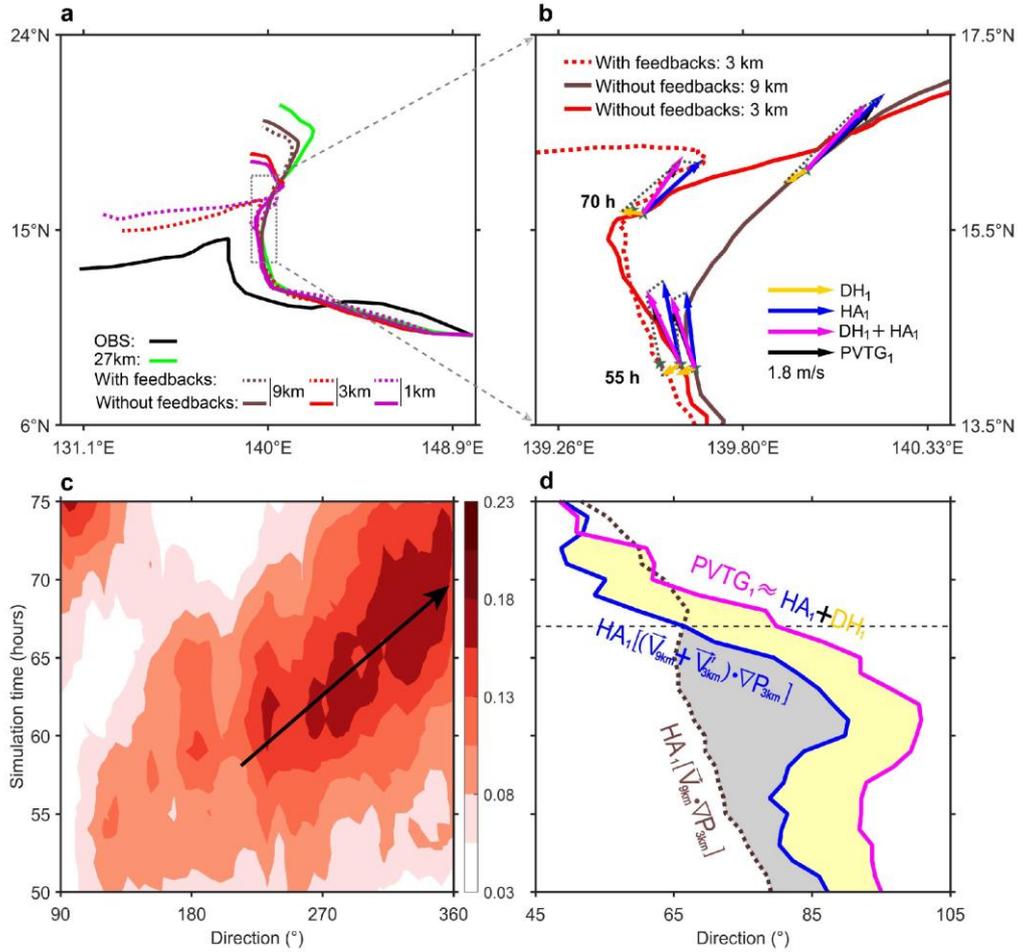

**Extended Data Figure 7 | Cumulonimbus clouds-generated asymmetric momentum anomalies triggering TFBs turning. a,** Simulated and observed tracks of Typhoon Kammuri in two-way (dotted) and one-way (solid) 9km- (brown), 3km- (green) and 1km-res (purple) downscalings, base 27km-res simulation (green), and the observation (black). **b,** TC moving vectors represented by wavenumber-1 potential vorticity tendency gradient[32] ($PVTG_1$) in 3km- and 9km-res one-way downscalings, zoomed-in from the black-dotted box in **a**. The $PVTG_1$ can be decomposed as PV horizontal advection ($HA_1$: blue) and cloud diabatic heating ($DH_1$: yellow), with asterisks marking TC positions at the instant. **c,** The 100-300 km radially-averaged rainwater mixing ratio (RMR) differences at azimuth-time section between 9km- and 3km-res simulations. The black arrow indicates cyclonic rotating of 3km-res RMR in typhoon spiral clouds relative to 9km-res RMR, after the 55$^{th}$ hour. **d,** Timeseries of angles of $PVTG_1$ (purple) and contributive vectors from $HA_1[(\mathbf{V}_{9km}+\mathbf{V'}_{3km}) \cdot \nabla P_{3km}]$ (blue) and $DH_1$ (yellow) in 3km-res one-way downscaling compared to $HA_1[\mathbf{V}_{9km} \cdot \nabla P_{3km}]$ ($HA_1$ contributive vector using 9km resolution mother-domain wind field) (brown). Here ($\mathbf{V}_{9km}+\mathbf{V'}_{3km}$) represents the 3km-res wind field downscaled from 9km-res mother-domain, which contains fine-scale asymmetric momentum anomalies (AMAs) generated by 3km-res resolved cumulonimbus

clouds. The difference between $HA_1[(V_{9km}+V'_{3km}) \cdot \triangledown P_{3km}]$ and $HA_1[V_{9km} \cdot \triangledown P_{3km}]$ (gray shaded) indicates the influence of AMAs on TC motion.

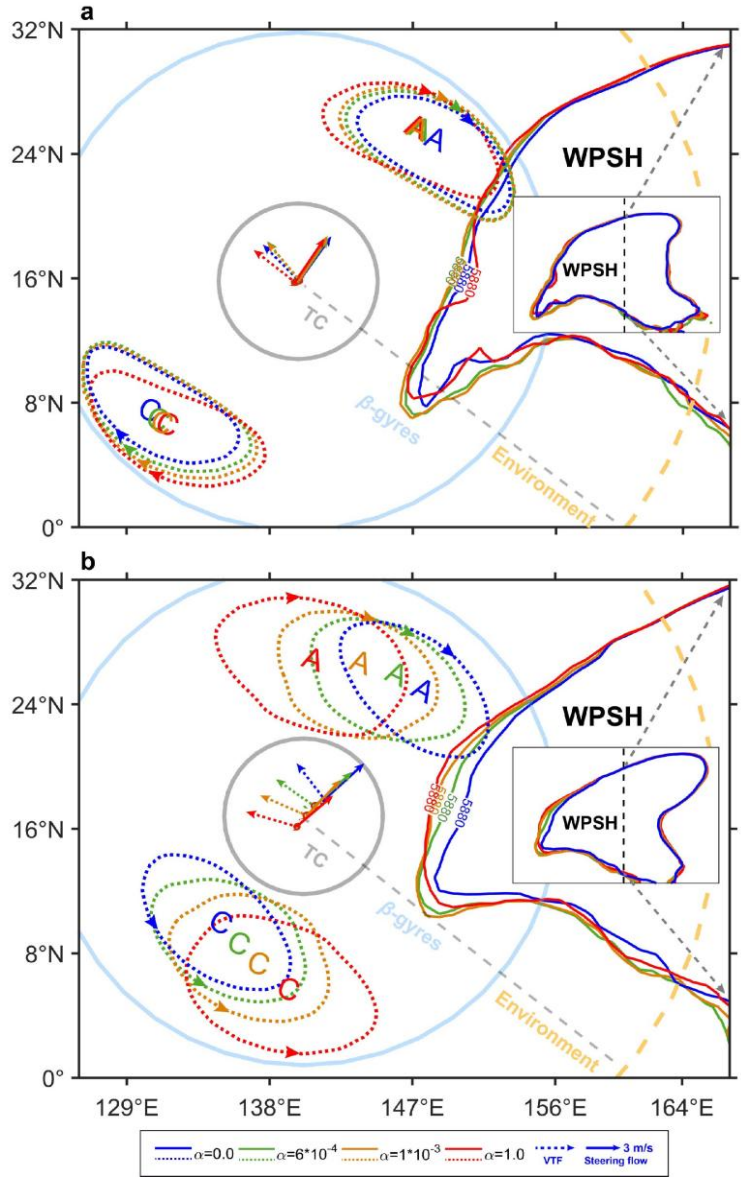

**Extended Data Figure 8 | Same as Fig. 4c but for the pre-turning and in-turning phases of Kammuri. a,** The TC β-gyres (color-dotted) starting being changed by fine-scale clouds at the early pre-turning (65$^{th}$-hour) point, and the WPSH environment (color-solid) being disturbed by the feedback of fine-scale clouds as shown by different feedback strengths [from no-feedback (α=0, blue) to full feedback (α=1, red)]. **b,** The TC β-gyres (color-dotted) having been reoriented by fine-scale clouds and their feedbacks to the environment at in-turning phase (80$^{th}$-hour), and the WPSH environment (color-solid) being adjusted due to the feedback of fine-scale clouds as shown by different feedback strengths [from no-feedback (α=0, blue) to full feedback (α=1, red)]. Solid and dotted arrows

indicate steering flows and VTFs respectively.